\documentclass[12pt]{article}

\usepackage{a4}
\usepackage{a4wide}

\usepackage{amsmath}
\usepackage{amscd}
\usepackage{amsfonts}
\usepackage{amssymb}
\usepackage{mathrsfs}
\usepackage{fancybox} 
\usepackage{enumerate}
\usepackage{ulem}
\usepackage{cite}
\usepackage{bm}
\usepackage{amscd}
\usepackage{graphicx}
\usepackage[dvips]{color}
\usepackage{epsfig}
\usepackage{subfigure}

\makeatletter

\@addtoreset{equation}{section}
\makeatother
\usepackage{amsthm} 

\def\const{\mathrm{const}}

\def\={\stackrel{\bullet}{=}}

\def\({\left(}
\def\){\right)}
\def\[{\left[}
\def\]{\right]}

\def\c\bar v{{\cal \bar v}}
\def\c\bar w{{\cal \bar w}}

\def\mbf{\mathbf}

\def \be {\begin{equation}}
\def \ee {\end{equation}}
\def \beqa {\begin{eqnarray}}
\def \eeqa {\end{eqnarray}}
\def \beal#1 {\begin{align}#1\end{align}}
\def \bes#1 {\begin{equation}\begin{split}#1\end{split}\end{equation}}
\def \nn {\notag\\}

\def\pmat#1{\begin{pmatrix}#1 \end{pmatrix} }

\makeatletter

\@addtoreset{equation}{section}
\makeatother

\begin{document}

\begin{titlepage}
\title{
\vspace{-2cm}
\begin{flushright}
\normalsize{ 
YITP-21-35 \\ 
}
\end{flushright}
       \vspace{1.5cm}
An analytic model for gravitational collapse of spherical matter under mixed pressure
\vspace{1.cm}
}
\author{
Shuichi Yokoyama\thanks{shuichi.yokoyama[at]yukawa.kyoto-u.ac.jp},\; 
\\[25pt] 
{\normalsize\it Center for Gravitational Physics,} \\
{\normalsize\it Yukawa Institute for Theoretical Physics, Kyoto University,}\\
{\normalsize\it Kitashirakawa Oiwake-cho, Sakyo-Ku, Kyoto 606-8502, Japan}
}

\date{}

\maketitle

\thispagestyle{empty}

\begin{abstract}
\vspace{0.3cm}
\normalsize
We investigate spherically symmetric gravitational collapse of thick matter shell without radiation in the Einstein gravity with cosmological constant. 
The orbit of the infalling thick matter is determined by imposing an equation of state for the matter near interface, where pressure constituted of the transverse component and the longitudinal one is proportional to energy density. 
We present analytic solutions for the equation of state and discuss parameter region to satisfy physical conditions such as the absence of the shell crossing singularity, the monotonic increase of the emergent infinite redshift surface and energy conditions.
We finally show that adopting the definition presented in arXiv:2005.13233 the total energy in this time-dependent system is invariant under the given time evolution. 

\end{abstract}
\end{titlepage}

\section{Introduction}
\label{Intro} 

The study of dynamical process for matter to collapse into a black hole is important not only in observational viewpoints but also in theoretical ones of general relativity. 
Oppenheimer and Snyder initiated to study a classical gravitational implosion of a compact star analytically \cite{1939PhRv...56..455O}, where homogeneous spherical pressureless gas falls into the center of the system without any radiation. (See also \cite{Datt1938}.) 
Subsequently gravitational collapses in various situations including the one with radiation were studied  \cite{Misner:1964je,Penrose:1964wq,1965gtgc.book.....H,1966ApJ...144..943V,1966ApJ...143..682M,10.1143/PTP.38.92}, which 
may incubate conceptually important ideas such as the cosmic censorship \cite{Penrose:1969pc} and the singularity theorem \cite{Penrose:1964wq}.
(See also \cite{Hawking:1969sw,Hawking:1973uf,Wald:1997wa,Senovilla_2015}.)

A traditional approach to the construction of an analytic solution to describe gravitational collapse is to prepare two solutions, one of which entails an event horizon in its spacetime and the other does not, and cut and glue them smoothly at their boundaries. 
Indeed the Oppenheimer-Snyder solution can be obtained \cite{Eardley:1978tr} by sewing two Tolman-Bondi space-times \cite{1934PNAS...20..169T,Bondi:1947fta}, whose metric has the trivial time-component but can describe the Schwarzschild space-time in Lemaitre coordinates \cite{1933ASSB...53...51L}. 
Such a sewing condition may be also referred to as the junction condition, which was investigated in several styles \cite{MSM1927,zbMATH03073849,lichnerowicz1955théories,Israel:1966rt}. 
(See also \cite{1981GReGr..13...29B}.)
Such a condition may be also recast in a modern form to request the second fundamental forms to match at the boundary,\footnote{The author received a comment from a referee that the analysis in \cite{1981GReGr..13...29B} is in fact incorrect, while the one in \cite{Israel:1966rt} is complete except for the case of null boundary and fits into the modern form.} which was used to construct analytic solutions to describe spherically symmetric gravitational collapse \cite{Fayos1991MatchingOT,PhysRevD.45.2732,PhysRevD.54.4862,Singh:1997iy} including cosmological constant  \cite{1991GReGr..23..471G,Cissoko:1998mx,Markovic:1999di,Lake:2000rm}. 
(See also \cite{Christodoulou:2008nj,Joshi:2011rlc} for references therein.)

Another approach is to find one good coordinate system to entirely cover the two space-times which are to be glued in order to describe the gravitational collapse \cite{Adler:2005vn}. 
This approach has advantages to circumvent technicality in the junction condition and to make it easy to get the whole picture, while it may be heuristic to find such a good coordinate system and the approach does not alway make the calculation simplified \cite{Misner:1974qy,Landau:1982dva}. However an adequate coordinate system also makes the calculation considerably simple, which will pay off enough for an effort to find. 

\begin{figure}[ht]
  \begin{center}
  \subfigure[Emergent black hole.]{
  \includegraphics[scale=.45]{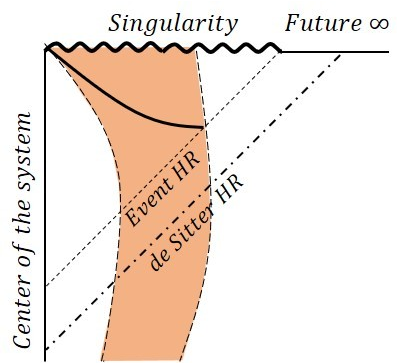}
  }
  \qquad
  \subfigure[Emergent local naked singularity 1.]{
  \includegraphics[scale=.45]{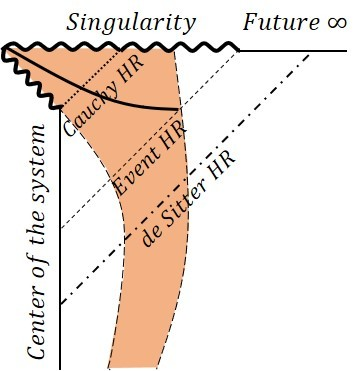}
  }
  \qquad
  \subfigure[Emergent local naked singularity 2.]{
  \includegraphics[scale=.45]{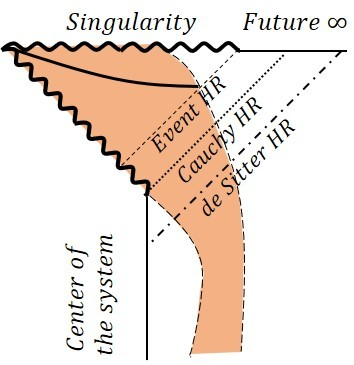}
  }
  \qquad
  \subfigure[Emergent global naked singularity.]{
  \includegraphics[scale=.45]{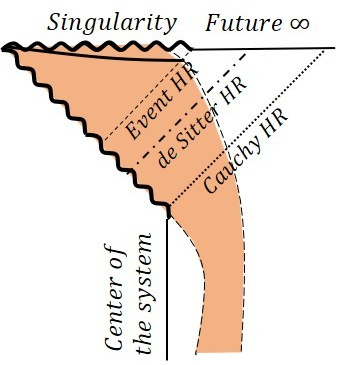}
  }
  \end{center}\vspace{-0.5cm}
  \caption{The possible local causal structures of the gravitational collapse of thick shell of fluid are depicted. ``HR'', ``Future $\infty$'' mean ``horizon'', ``Future infinity'', respectively. The space-time is separated by the thick shell of fluid, whose region is filled with some color. We consider a situation where there exists a positive cosmological constant or dark energy, which generates de Sitter horizon and modifies the far asymptotic structure from the flat case investigated in \cite{Eardley:1978tr}. 
  The bold line extending from the singularity to the event horizon is the infinite redshift surface or so-called the trapping horizon.}
  \label{PC}
\end{figure}

A main purpose of this paper is to take the second approach and extend results of spherically symmetric gravitational collapse of light-like shell in \cite{Adler:2005vn} to that of general fluid including cosmological constant. 
(See also \cite{Alberghi:1998xe,Alberghi:2006zy,Tippett:2011hz}.)
In fact, it is not trivial at all to construct such a solution, since it was shown in earlier studies that an emergent curvature singularity is not generally covered by an event horizon and may become naked \cite{Eardley:1978tr,Christodoulou:1984mz,PhysRevD.43.1416,Shapiro:1991zza}. (See also \cite{cmp/1104115427,Christodoulou:1987vu,Christodoulou:1987vv,Dwivedi:1989pt,Christodoulou:1991yfa,PhysRevD.45.2147,Shapiro:1992heg,Christodoulou:1994hg}.)
We summarize the possible causal structures of gravitational collapse by the local Penrose-Carter diagrams in Fig.\ref{PC}.
The violation of the cosmic censorship implies the breakdown of the physics law and a solution with naked singularity emergent is not physically favored \cite{Hawking:1976ra}.
Therefore it is meaningful and useful to construct an analytic solution which describes gravitational collapse without any naked singularity even though it is only a toy model. 

In this paper we avoid an unphysical solution with naked singularity by choosing a good coordinate system as an ansatz for the metric at the beginning. 
Then we impose several physical conditions on the solution. One is such that there does not happen any shell crossing singularity \cite{Eardley:1978tr}. 
Indeed upon this extension the absence of the shell crossing singularity becomes a non-trivial obstruction, which constrains the parameter space of consistent solutions. (See also \cite{Joshi:2012ak}.)
Another is the infinite redshift surface monotonically increase, which will turn out to constrain the gradient of the orbit of infalling fluid from the above. The others are energy conditions such that the local energy density is non-negative and the energy flux vector is timelike outside the emergent event horizon.
Another characteristic point of our extension is to consider gravitational collapse of general fluid under mixed pressure of both the transverse component and the longitudinal one in its equation of state. 
We can still solve such a gravitational collapsing system with fluid under such a general pressure analytically.

Furthermore we compute the total energy of the system adopting the definition of charges presented in \cite{Aoki:2020prb}. (See also \cite{Fock:1959}.) The proposed definition of charges is done so as to extend the one in the flat space-time to a general curved space-time with manifest invariance under general coordinate transformation. This definition provides a simple way to test whether a charge is conserved or not by checking the associated vector field for the charge to satisfy an equation referred to as the conservation equation \cite{Aoki:2020nzm}. We show that the time evolution vector field in the matter collapsing system studied in this paper satisfies the conservation condition, and confirm that the total energy is indeed invariant under the given time evolution by explicit computation.    

The rest of this paper is organized as follows. 
In section \ref{Model} we analyze a model of gravitational collapse of spherical thick matter shell, which outermost shell obeys an equation of state with mixed pressure of the transverse component and the longitudinal one, and solve it to satisfy the physical conditions. 
In section \ref{Conservation} we compute the total energy flow of this system, and show that the time evolution vector field satisfies the conservation condition and that the total energy is invariant under the given time evolution. Section \ref{Discussion} is devoted to summary and discussion.

\section{Model of gravitational collapse of thick matter shell} 
\label{Model}

In this section we fix the setup to study the formation of a spherically symmetric black hole in a $d$-dimensional spacetime with a general value of cosmological constant $\Lambda$, where $d$ is greater than two.%
\footnote{
Our primary interest is in four dimensional system, but performing analysis in general dimensions we can also extract some interesting information at $d=3$ as seen below. } 
For this purpose, 
due to the ignorance of what types of matter with an initial configuration collapses catastrophically into a black hole, we take a  strategy to first look for a good coordinate system and a metric which can describe a gravitational collapse of our interest and set it as an ansatz. A desired metric describes a situation such that there exists fluid infalling into the origin at early times and only a static black hole exists in the whole space at late times. As such a coordinate system, following \cite{Adler:2005vn}, in which a gravitational collapse of null dust was studied, we adopt the original form of the Eddington-Finkelstein coordinate system:%
\footnote{ 
The Eddington-Finkelstein coordinate system is not fully general to describe a time-dependent spherically symmetric system. For example, it does not contain the Tolman-Bondi spacetime, which has two unknown functions: $ds_{\rm TB}^2=-dt^2 +X(t,r)dr^2 + Y(t,r)\tilde g_{ij} dx^i dx^j$. However, the original form of the Eddington-Finkelstein coordinates suits to describe spherical matter collapse without any radiation, while other coordinate systems such as the Vaidya one may not. 
}
\be
ds^2 =- (1+u) dt^2 - 2u dt dr+(1-u)dr^2 + r^2 \tilde g_{ij} dx^i dx^j. 
\label{EF}
\ee
Here $\tilde g_{ij}$ is the metric for a $d-2$ dimensional Einstein manifold whose Ricci tensor is $\tilde R_{ij} = (d-3)\tilde g_{ij}$, and $u$ is a function of $t$ and $r$ such that 
\be
u(t,r) = -\frac{2\Lambda r^2}{(d-2)(d-1)}- \frac{m(t,r)}{r^{d-3}} ,
\label{u}
\ee
where $m(t,r)$ is a matter profile function which could describe the dynamics of formation or deformation of a spherically symmetric black hole. 
Indeed, the metric \eqref{EF} with vanishing $m(t,r)$ describes the AdS or dS space-time for $\Lambda$ negative or positive, respectively, while for positive constant $m(t,r)$ it describes a static spherically symmetric black hole. This can be easily seen by transforming \eqref{EF} into the usual Schwarzschild metric when $u$ is a function only of $r$:
\beal{
ds^2=& - (1+u(r)) (dx^0)^2 +{1\over 1+u(r)} dr^2 + r^2\tilde g_{ij}dx^idx^j , 
}
where $x^0 =t + \int dr { u(r) \over 1+u(r)}$. 
Thus the event horizon emerges if $m(t,r)>0$, while another event horizon  called the de-Sitter or cosmological event horizon will also appear when $\Lambda>0$ \cite{Carter:1970ea,Carter:1973rla}. Note that in the original form of the Eddington-Finkelstein coordinates the event horizon and the infinite redshift surface are identical. 

The time direction in the metric \eqref{EF} can be freely chosen. Here we determine it so that the infalling null radial geodesic does not depend on the function $u$, so that it becomes the same as in the flat spacetime, $t=-r +\const$.  Then the outgoing radial light ray is $dr/dt = (1+u)/(1-u)$. This suggests that there is an infinite redshift at $u=-1$. Note that any light does not come out to the spacial infinity if the cosmological constant is nonzero. 

We wish to study whether there exists any matter which collapses gravitationally in the manner described by \eqref{EF}. 
To this end, assuming the metric \eqref{EF} as the ansatz for the Einstein equation, $ R^\mu\!_\nu -\delta^\mu\!_\nu\frac{R}{2} + \delta^\mu\!_\nu \Lambda =8\pi G_N T^\mu\!_\nu ,$ we determine the form of the matter energy momentum tensor:
\bes{ 
T^t\!_t =& 
{d-2 \over 16\pi G_N}
{\partial_r  (r^{d-3} \delta u )\over r^{d-2}},  \\
T^t\!_r =& 
{d-2 \over 16\pi G_N}{\partial_t u \over r} = - T^r\!_t, \\
T^r\!_r =& 
{d-2 \over 16\pi G_N} ({\partial_r  (r^{d-3} \delta u )\over r^{d-2}} -2 { \partial_t u \over r}), \\
T^{i}\!_{j}
=&{\delta^i\!_j\over 16\pi G_N} {(\partial_r  - \partial_t)^2(r^{d-3}\delta u)\over r^{d-3}},
\label{EMT}
}
where $G_N$ is the Newton constant, and $\delta u( t,r) = -\frac{m(t,r)}{r^{d-3}} $.\footnote{ 
One might wonder whether the strategy here to determine the energy momentum tensor from a given metric configuration is valid or meaningful. 
The Einstein equation can only determine the form of the energy momentum tensor and is not sufficient to specify the matter. 
Taking advantage of this we perform a general analysis of gravitational collapse of unspecified matter or fluid as seen below. A goal is to elucidate some important property of gravitational collapse of fluid from this general analysis so as to extract a general lesson for further analysis. 
}
Remark that the $\Lambda$ dependence cancels due to the fine-tuned coefficient of $\Lambda$ in \eqref{u}.\footnote{ We have checked that \eqref{EMT} satisfies the covariant conservation $\nabla_\mu T^\mu\!_\nu=0$ in four dimensions.} 
In order to justify the ansatz of the metric \eqref{EF} as a result of the physical process of infalling matter, 
we need to show a local frame well-defined during the process which enables us to read off physically meaningful quantities relative to it. From such a local observer, the stress energy tensor is related to macroscopic quantities by \cite{1935ApJ....82..435H} 
\be 
T^\mu\!_\nu= \rho v^\mu v_\nu +q^\mu v_\nu + v^\mu q_\nu + P^\mu\!_\nu,
\ee 
where $v_\mu$ is the velocity of the infalling fluid, $\rho$ is the energy density, $q^\mu$ is a so-called heat flux and $P^\mu\!_\nu$ is a stress tensor perpendicular to the fluid velocity.
In a timelike frame, in which the fluid velocity satisfies $v^\mu v_\mu=-1$, these quantities are computed as 
\beal{ 
\rho=& v_\mu T^\mu\!_\nu v^\nu , ~~
q^\mu = -h^\mu\!_\rho T^\rho\!_\nu v^\nu, ~~ 
P^\mu\!_\nu = h^\mu\!_\rho T^\rho\!_\sigma h^\sigma\!_\nu
}
where $h^\mu\!_\nu= \delta^\mu\!_\nu + v^\mu v_\nu$ is the projector. In the current spherically symmetric system, the fluid velocity only has the time and radial components, $v^\mu=(v^t,v^r,\vec0)$, and the stress tensor can be written as 
\be 
P^a\!_b = p_{\perp} h^a\!_b, ~~~ 
P^i\!_j = p_{\parallel} \delta^i\!_j 
\ee
where $a,b=t,r$. 
The components of the fluid velocity are constrained to satisfy $v^\mu v_\mu=-1$, which is solved as 
\be 
v^r = \frac{uv^t \pm \sqrt{-1+u +(v^t)^2 } }{1-u}. 
\ee
In order for a local frame to be well-defined for an arbitrary function of $u(t,r)$, there has to disappear the square root as well as the denominator in the above expression. 
This can be realized only if the time component of the fluid velocity is given by $v^t = \pm (1-\frac{1}2u) $. 
Assuming the fluid vector to satisfy $v^\mu=(1,0,\cdots,0)$ for the case $u=0$, the fluid velocity to satisfy these constraint is uniquely determined as 
\be 
v^\mu = \pmat{1-\frac{1}2u \\ \frac {1}2u \\ \vec0}. 
\ee
Note that 
\be 
v^\nu \nabla_\nu v^\mu = \frac{\partial_ru -\partial_t u}4 \pmat{-u \\ 2+u \\ \vec0}.   
\ee
Then the observable quantities relative to this physics-based frame are computed as follows. 
\bes{
\rho=&  {d-2\over 16\pi G_N }\frac{\partial_r m(t,r) }{r^{d-2}},
\\
q^\mu =&\frac{d-2}{32\pi G_N}\frac{\partial_t m(t,r)}{r^{d-2}}\pmat{u \\ -(2+u) \\ \vec0  }, \\
p_{\perp}
=&{d-2\over 16\pi G_N }(\frac{-\partial_r m(t,r)}{r^{d-2}} +2{\partial_t m(t,r)\over r^{d-2}}),
\\
p_{\parallel}=&{1\over 16\pi G_N } 
\frac{(\partial_r -\partial_{t})^2(-m(t,r)) }{r^{d-3}}. 
\label{densityPressure}
}
In particular, the energy density per unit volume read off in this way reproduces the expected total energy of the system, as we shall see below. 
Note that in order for this energy density to be non-negative in the whole region, the parametric function $m(t,r)$ is constrained to satisfy 
\be 
\partial_r m(t,r)\geq0.
\label{rhoPositive}
\ee 

Now we assume the matter profile to form shell structure, so that it is a function of the one dimensional subspace such that 
\be
m(t,r) = m \theta(r) F\left(\dfrac{t+h(r)}{\Delta}\right).
\label{matterProfile}
\ee
Here $m$ is a positive constant related to the black hole mass at the end, $h(r)$ is a function to describe an orbit of a piece of matter consisting of the shell. 
We insert the step function $\theta(r)$, which plays no role to construct the solution in this section but a technically important role to compute the contribution of the energy of emergent curvature singularity in the next section.
$F(x)$ is an ``upslope'' function: it vanishes from $x=-\infty$ to the origin, at which it starts to increase monotonically from $0$ to $1$ up to $x=1$ and keeps the same value afterwards from $x=1$ to $x=\infty$.%
\footnote{On the other hand, in the case to study the matter explosion, a ``downslope'' function is suitable, which is described by $F(1-x)$. Such a solution to describe the process of black hole diffusion will be obtained by flipping the time direction for the solutions obtained below. }
In this normalization of the upslope function, a positive constant $\Delta$ represents the thickness of the shell. 
The upslope function plays a role of the continuous label of the matter shell such that the concentric slice of matter shell labeled by $F=\lambda \; (0\leq\lambda\leq1)$ has the world-line $t=-h(r) + F^{-1}(\lambda) \Delta$, where $F^{-1}$ is the inverse function of $F$ only in the region $[0,1]$ so that $F^{-1}(0)=0, F^{-1}(1)=1$. In particular, $F(0)=0$ labels the innermost shell while $F(1)=1$ labels the outermost one.
Note that a discrete sequence of thin matter shells can be realized by choosing the upslope function as a multi-step function. 
This upslope function is used as a physical shell attribute in this system as we shall see in section \ref{Conservation}.
In what follows we assume $F'(0)=F'(1)=0$ for the continuity of the density and the radial component of pressure, and set $h(0)=0$ using the time shift.

Our goal is to determine the form of the function $h(r)$ for given parameters $\Delta, m$ and $F(x)$.
We do this for the matter near the interface specified by $F=1-\epsilon$ with $\epsilon$ a small parameter to satisfy an equation of state such that 
\be 
P = w\rho,
\label{EOS}
\ee
where $w$ is a constant, $\rho$ is the density, and $P$ is the pressure the matter at the interface receives. 
As seen from \eqref{densityPressure} the pressure in the equation of state can be generally composed of the radial component and the angular one. 
In the following subsections, we determine $h(r)$ in a couple of cases for $P$ to be composed only of the angular component, only of the radial component, and of their mixture in order. 

Since we impose a condition only at the junction, it is not guaranteed that a solution obtained in this way is consistent in the whole space-time region. We require a solution to be free from a shell crossing singularity \cite{Eardley:1978tr}. 
The absence of such a shell crossing singularity restricts the parameter region. 
In order to avoid a shell crossing singularity at the junction we impose a condition for $h(r)$ to increase monotonically. If this is not satisfied, then the shell ansatz \eqref{matterProfile} breaks down in some space-time region.
Note that this condition also satisfies the non-negative energy density condition \eqref{rhoPositive} in the current setup. Indeed in the shell ansatz \eqref{matterProfile}, the condition \eqref{rhoPositive} reduces to 
\be 
F'\left(\dfrac{t+h(r)}{\Delta}\right) \frac{h'(r)}\Delta \geq 0.
\label{rhoPositive1}
\ee 
Since we chose the function $F$ to be a non-trivial upslope function and $\Delta$ positive, the condition \eqref{rhoPositive1} is satisfied if $h'(r)\geq0$. 
In addition we also impose a condition that the emergent infinite redshift surface monotonically increases. 
Note that for positive cosmological constant the smaller event horizon is the black hole one, whose radius is increasing, while the bigger one is the cosmological event horizon, whose radius is decreasing during the matter collapsing into a black hole \cite{Gibbons:1977mu}.
The infinite redshift surface is given by an equation $u=-1$, where $u$ is given by \eqref{u}, since the outgoing radial null geodesic is determined from $dr/dt = (1+u)/(1-u)$ as shown above. Combining with \eqref{matterProfile} the emergent infinite redshift surface can be explicitly written as 
\be 
t=-h(r) +\Delta F^{-1}\(\frac{r^{d-3}}m(1-\frac{2\Lambda}{(d-2)(d-1)}r^2 )\)
\ee
up to when the surface reach the final black hole horizon radius $r=r_{BH}$, which is determined from $\frac{r_{BH}^{d-3}}m(1-\frac{2\Lambda}{(d-2)(d-1)}r_{BH}^2 )=1$. 
The absence of a shell crossing singularity requires the right hand side to increase monotonically with respect to $r$. 

The two conditions can be summarized as 
\be 
0< h'(r) <  \frac{\Delta }mr^{d-4} (d-3 -\frac{2\Lambda}{d-2}r^2 ) (F^{-1})'\(\frac{r^{d-3}}m(1-\frac{2\Lambda}{(d-2)(d-1)}r^2 )\)
\label{NoCrush}
\ee
for $r<r_{BH}$. 
We have comments on the second inequality in \eqref{NoCrush}, which comes from the second condition above. 
The second inequality is sensitive to the signature of the cosmological constant. 
If the cosmological constant is not positive, then the right-hand side can be arbitrary large by taking $\Delta$ to be large for any upslope function $F$. This means that the second inequality in \eqref{NoCrush} is satisfied if $\Delta$ is sufficiently large when $\Lambda\leq0$. 
On the other hand, a careful analysis is necessary for the case of a positive cosmological constant. 
It is interesting to comment at three dimensions. 
At $d=3$, \eqref{NoCrush} reduces to 
\be 
0< h'(r) <  \frac\Delta m(-2\Lambda r ) (F^{-1})'\(\frac{1}m(1-\Lambda r^2 )\).
\ee
This inequality cannot be satisfied unless the cosmological constant is negative. This shows that a (spherically symmetric) black hole in three dimensions cannot form unless there is no negative cosmological constant in the Einstein gravity \cite{Banados:1992wn,Ida:2000jh}. 

In addition to the above conditions, in order for infalling general fluid to be physical, its energy flux vector should be causal outside the event horizon.\footnote{The argument of the flux energy condition and the discussion in appendix \ref{EnergyCondition} were added to answer a question asked by a referee. The author would like to thank the referee for pointing this out. }
Since the energy flux vector is given by $-T^\mu\!_\nu v^\nu$, the condition is described as 
\be 
(\partial_t m(t,r))^2\leq(\partial_r m(t,r))^2.
\ee
In the shell ansatz \eqref{matterProfile} and the choice of $F$ as an upslope function, this condition reduces to 
\be 
1\leq h'(r).
\label{FEC0}
\ee
for $r_{EH}\leq r$ with $r_{EH}$ the radius of the emergent event horizon. 
We comment on the constraint from standard energy conditions in appendix \ref{EnergyCondition}. 

Below we determine the form of $h(r)$, and subsequently study a qualitative parameter region to satisfy \eqref{NoCrush} implied by numerical calculation. 

\subsection{Transverse pressure} 
\label{Transverse}

We first consider a case where the pressure in the equation of state is composed only of the angular component, so that the equation of state is $p_{\parallel}=w_\parallel^0\rho$. 
Plugging this and \eqref{densityPressure} into \eqref{EOS} we find
\be
y (h'(r)-1)^2   +  h''(r)  = - \frac {w_\parallel}r h'(r),
\label{angular}
\ee
where $w_\parallel=(d-2)w_\parallel^0, ~ y= {F''(1-\epsilon) \over F'(1-\epsilon)}\Delta$. We rewrite \eqref{angular} in more useful form later as   
\be
y w_\parallel H(r)^2   +  H'(r)  = - \frac {w_\parallel}r H(r) - \frac1r,
\label{angularH}
\ee
where 
\be 
H(r) = \frac{h'(r)-1}{w_\parallel}. 
\ee

For $y=0$ this differential equation can be easily solved as $h(r) = \frac{c_0}{1-{w_\parallel}} r^{1-{w_\parallel}}$, where we used $h(0)=0$ and $c_0$ is an integration constant, unless ${w_\parallel}=1$. A solution with ${w_\parallel}=1$ is obtained by formally taking the limit of this solution and ignoring a constant term: $h(r) = c_0 \log r$. 
We fix the integration constant $c_0$ by requesting that the shell of the pressureless matter of ${w_\parallel}=0$ is the light-like shell or null-dust. This is satisfied if $c_0=1$ at ${w_\parallel}=0$. For simple presentation we set $c_0=1$ in what follows:
\be 
h(r) = \frac{r^{1-{w_\parallel}}}{1-{w_\parallel}}.
\label{yvanish}
\ee
This implies that $w_\parallel$ has to be smaller than unity for a reasonable solution.

For $y>0$, it is difficult to solve \eqref{angular} analytically in terms of $h(r)$, while it is possible in terms of $h'(r)$ or to solve  \eqref{angularH} as follows. 
\beal{
H(r) 
=&\frac{ \Gamma (2-w_\parallel) J_{-w_\parallel}\left(2 \sqrt{r y w_\parallel}\right)-c_y \Gamma (w_\parallel) J_{w_\parallel}\left(2 \sqrt{r y w_\parallel}\right) }{\sqrt{r y w_\parallel} \left(\Gamma (2-w_\parallel) J_{1-w_\parallel}\left(2 \sqrt{r y w_\parallel}\right) + c_y \Gamma (w_\parallel) J_{w_\parallel-1}\left(2 \sqrt{r y w_\parallel}\right)\right)} ,
\label{angularH1}
}
where $c_y$ is an integration constant and $J_\nu(z)$ is the Bessel function of the first kind.  
Note that in order to reach this form we used the following recurrence equation for the Bessel function 
\be 
J_{\pm(w-2)}(z) = \frac{\pm2 (w-1)}{z}J_{\pm(w-1)}(z) -J_{\pm w}(z).
\ee
We fix the integration constant $c_y$ for this solution to reduces to the one \eqref{yvanish} by taking the limit $y\to0$. 
To this end we expand the solution \eqref{angularPsol} at $y\sim0$ keeping $|{w_\parallel}| \ll 1$. 
Employing the asymptotic expansion of the Bessel function $J_\alpha(z) \sim \frac1{\Gamma(\alpha+1)}({z\over2})^\alpha $ around $z\sim0$, we find 
\beal{
H'(r) =\frac1{w_\parallel}- \frac{(1-{w_\parallel})  (r y w_\parallel)^{-{w_\parallel}}}{c_y} + \cdots ,
}
where the ellipsis contains higher ordered terms.
Therefore it is sufficient to choose as 
\be 
c_y ={1-{w_\parallel} \over y^{w_\parallel} {w_\parallel}^{{w_\parallel}-1}}
\label{cy}
\ee
for the reduction to \eqref{yvanish} with $y\to0$. 
As a result we obtain   
\be
H(r) = \frac1{\sqrt{r y w_\parallel}}\frac{(y w_\parallel)^{w_\parallel} \Gamma (2-{w_\parallel}) J_{-{w_\parallel}}\left(2 \sqrt{r y w_\parallel}\right)+({w_\parallel}-1) {w_\parallel} \Gamma ({w_\parallel}) J_{w_\parallel}\left(2 \sqrt{r y w_\parallel}\right)}{(y w_\parallel)^{w_\parallel} \Gamma (2-{w_\parallel}) J_{1-{w_\parallel}}\left(2 \sqrt{r y w_\parallel}\right) - ({w_\parallel}-1) {w_\parallel} \Gamma ({w_\parallel}) J_{{w_\parallel}-1}\left(2 \sqrt{r y w_\parallel}\right)}. 
\label{angularPsol}
\ee
Note that we could not further perform the integration analytically to obtain $h(r)$.

\begin{figure}
  \begin{center}
  \subfigure[$w_\parallel=0.1, y=0.1$.]{
  	\includegraphics[scale=.45]{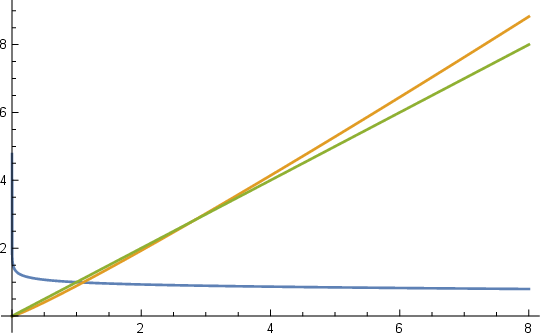}
  }
  \qquad
  \subfigure[$w_\parallel=0.1, y=1.2$.]{
  	\includegraphics[scale=.45]{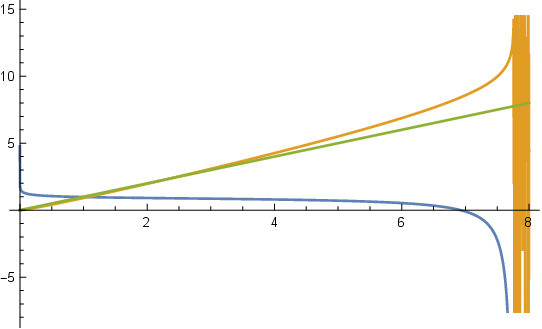}
  }
  \qquad
  \subfigure[$w_\parallel=0.95, y=0.2$.]{
  	\includegraphics[scale=.45]{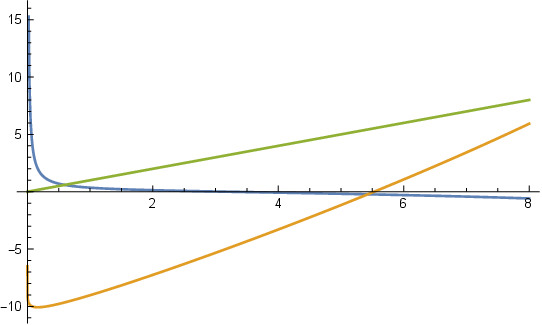}
  }
  \end{center}\vspace{-0.5cm}
  \caption{The blue, yellow and green curves depict the functions $h'(r),h(r),r$, respectively, where the horizontal axis is the $r$ coordinate, with parameters chosen as shown. }\label{wy}
\end{figure}

Let us study a region where $h(r)$ is a monotonically increasing function or $h'(r)$ is positive. It can be confirmed that the condition is always satisfied if both $y$ and $w_\parallel$ are sufficiently small by using the asymptotic formula of the Bessel function of the first kind $J_\nu(z) \sim \sqrt{\frac2{\pi|z|}} \cos(z-\frac\pi4-\frac{\nu\pi}2)$ for $z\sim\infty$.
However, numerical results suggest that the solution does not meet the condition and hit a crossing singularity from a certain $r$ if $y$ or $w_\parallel$ is close to one as seen from Figure~\ref{wy}. 
In fact the numerical solution for $h(r)$ starts oscillating at large $r$ and is not reliable in this regime. 

We shall give a comment on the second inequality in \eqref{NoCrush} in the next section drawing a graph with parameters suitably chosen. 

\subsection{Longitudinal pressure }
\label{Longitudinal}

Next we consider a case where the pressure in \eqref{EOS} is given by the radial component, so that $p_{\perp}=w_\perp\rho$. It seems that this case was not studied before, but it is theoretically possible and becomes a basis for the study of mixed pressure in the next subsection. 
Substituting \eqref{densityPressure} into \eqref{EOS} we obtain 
\be 
h'(r) ={ 2 \over 1+w_\perp}  ~~ \rightsquigarrow ~~  
h(r) ={2 \over 1+w_\perp}r. 
\label{radialPsol}
\ee

As a result the collapse of matter shell to satisfy $p_{\perp}=w_{\perp}\rho$ outermost is equivalent to the one to satisfy $p_{\parallel}=0$ with $F''(1)=0$ and the integration constant $c_0={2 \over 1+w_\perp}$.

\subsection{Mixed pressure }
\label{Mixed}

At the end of this section we consider the collapse of matter under mixed pressure such that $P=p_{\parallel}+v p_{\perp}$, where $v$ is a non-negative parameter. 
Employing \eqref{densityPressure} the equation of state \eqref{EOS} can be computed as 
\beal{
y (h'(r)-1)^2   +  h''(r) - \frac{\bar v}r = -\frac{\bar w}r h'(r),
}
where $\bar v=2(d-2) v, \bar w=(d-2)(v+w)$. 
This can be further rewritten as 
\be
y (\bar w-\bar v) H(r)^2   +  H'(r)  = - \frac {\bar w}r H(r) - \frac1r,
\label{mixedH}
\ee
where 
\be 
H(r) = \frac{h'(r)-1}{\bar w-\bar v}. 
\ee
This differential equation \eqref{mixedH} in this form can be obtained from the previous one \eqref{angularH} by replacing $y w_\parallel \to y (\bar w-\bar v) $ and $w_\parallel \to \bar w$. 
Therefore the solution for \eqref{mixedH} can be also obtained from that for \eqref{angularH}, which is given by \eqref{angularPsol}, by the same replacement:
\beal{
\frac{h'(r)-1}{\bar w-\bar v} 
=& \frac{\Gamma (2-\bar w) J_{-\bar w}\left(2 x\right)-c \Gamma (\bar w) J_{\bar w}\left(2 x\right)}{x \left(\Gamma (2-\bar w) J_{1-\bar w}\left(2 x\right) +c \Gamma (\bar w) J_{\bar w-1}\left(2 x\right)\right)},
\label{h'MixedPressure}
}
where $c$ is an integration constant, $x= \sqrt{r y(\bar w-\bar v)}$. 
We fix the integration constant $c$ by requesting this solution to reduce to \eqref{angularPsol} by taking the limit $\bar v\to+0$. 
This can be easily fixed by the above replacement of parameters for \eqref{cy} taking into account $\bar v\sim0$: 
\be 
c = \frac{\bar w(1-\bar w) }{(y\bar w)^{\bar w}} + \alpha \bar v,  
\ee
where $\alpha$ is an arbitrary number which can be dependent on ${\bar v}$ and ${\bar w}$ in a non-singular manner.  
We fix $\alpha$ so that this solution reduces to the one with only radial pressure, \eqref{radialPsol}, in the limit ${\bar v},{\bar w}\to\infty$ holding their ratio fixed.
For simplicity we take the limit $\bar v,\bar w\to\infty$ as well as $y\to0$ with $y (\bar w-\bar v)$ fixed. Then $x$ becomes constant in this limit. 
If we choose $\alpha\not=0$, the leading term of the above solution \eqref{h'MixedPressure} can be computed as 
\beal{
h'(r) -1 \sim& \frac{ (\bar w-\bar v) \left(- J_{\bar w}\left(2 x\right)\right)}{x J_{{\bar w}-1}\left(2 x\right)}
\sim -1 + \frac {\bar v}{\bar w}, 
}
where we used the asymptotic formula $J_\nu(z) \sim \frac1{\sqrt{2\pi\nu}}(\frac{ez}{2\nu})^\nu $ with $\nu \sim \infty$. 
Thus any nonzero $\alpha$ gives a desired solution. 
For $\alpha=1$, the final solution is expressed as 
\beal{
h'(r)
=1 + \frac{(\bar w-\bar v)[(y\bar w)^{\bar w}\Gamma (2-\bar w) J_{-\bar w}\left(2 x\right)-\{\bar w(1-\bar w)  + (y\bar w)^{\bar w} \bar v\} \Gamma (\bar w) J_{\bar w}\left(2 x\right)]}{x \left[(y\bar w)^{\bar w} \Gamma (2-\bar w) J_{1-\bar w}\left(2 x\right) +\{\bar w(1-\bar w)  + (y\bar w)^{\bar w} \bar v\} \Gamma (\bar w) J_{\bar w-1}\left(2 x\right)\right]}.
}

Let us comment on the valid parameter region where $h(r)$ is a monotonically increasing function or $h'(r)$ is positive including the parameter $\bar v$. Numerical results suggest that $h'(r)>0$ for sufficiently small $\bar v$ but not with $\bar v$ closed to one keeping both $y$ and $\bar w$ sufficiently small. If one of $y, \bar w, \bar v$ becomes close to one, we observe that the solution hits a shell crossing singularity at some $r$. 
To summarize, the valid parameter regime is (i) $\bar v, |\bar w|, y \ll 1$ with $\bar v y >0$, or (ii) $|\bar v|, |\bar w| \gg1 $ with $\bar v y >0, y \ll 1, |\bar v/\bar w| \sim1$.

\section{Conservation of energy}
\label{Conservation} 

In this section we investigate the flow of energy and energy density in the system studied in the previous section. 

Since all the matter collapses into a black hole in the process, the total energy is expected to be conserved.
It was shown in \cite{Herrera:1997plx} that the Misner-Sharp mass \cite{Misner:1964je} in similar gravitational collapsing systems is invariant under the co-moving time differentiation \cite{Bondi:1947fta}.
On the other hand, there has been a proposal of a manifestly covariant definition of the total energy in field theory on a general curved space-time \cite{Aoki:2020prb} 
\beal{
E=& \int_{\mbf R^{d-1}}\! d^{d-1} x\sqrt {|g|} T^t\!_\mu n^\mu,
\label{Energy} 
}
where $n^\mu$ is the time evolution vector field given by $n^\mu=-\delta^\mu_t$. 
In what follows we confirm that the energy defined by \eqref{Energy} is invariant under the usual time evolution.  

An advantage in the definition \eqref{Energy} is that whether the charge of the form \eqref{Energy} conserves or not can be easily determined by whether the vector field to define the charge satisfies the conservation condition \cite{Aoki:2020nzm}
\be 
T^\nu\!_\mu \nabla_\nu n^\mu = 0. 
\label{conservationTest}
\ee
In the current situation one can easily check that this equation is satisfied using \eqref{EMT}. Thus we conclude that the charge $E$ defined by \eqref{Energy} is conserved for all the cases studied in section \ref{Model} as long as the solution is meaningful without any shell crossing singularity. 

More explicitly we compute the total energy separating the time periods: $\{t<0\}$, when the thick matter shell is on-going falling before the black hole forms, $\{0\leq t < \Delta\}$, the black hole is forming, and $\{\Delta \leq t\}$, the thick matter shell completes infalling. 
Plugging \eqref{EMT} into \eqref{Energy} we obtain 
\beal{
E=&{d-2 \over 16\pi G_N} V_{d-2} \int_0^\infty dr  {\partial_r  (m(t,r) )} ,
\label{Energy2}
}
where $V_{d-2} = \int d^{d-2}x \sqrt{\tilde g}$ is the volume of the internal Einstein manifold.

When $t<0$, the innermost shell specified by $F=0$ does not reach the origin: $ 0<h^{-1}(-t)$.
Therefore we can compute the total energy as 
\beal{
E=&{d-2 \over 16\pi G_N} V_{d-2}m (F(1) - F(0)) = M, 
}
where $M={d-2 \over 16\pi G_N} V_{d-2} m$. 

At $t=0$ the black hole starts to form, and it becomes growing up to $ t=\Delta$. In this period both the infalling matter shell and the black hole exist. We denote the energy of the former by $M_1$ and that of the latter by $M_2$.
Then the total energy \eqref{Energy2} is given by the sum of two: $E=M_1+M_2$. 
$M_1$ is obtained by acting the differentiation with respect to $r$ on the upslope function $F$ in $m(t,r)$.
Thus it is computed as 
\beal{
M_1
=&  {d-2 \over 16\pi G_N} V_{d-2}  \int_0^\infty dr  \partial_r F(\frac{t + h(r)}\Delta) 
= M(1 - F(\frac t\Delta )), 
}
where we used $h(0)=0$ and the assumption that $h(r)$ increases monotonically to infinity.
On the other hand, 
$M_2$ is obtained by acting the differentiation with respect to $r$ on the step function $\theta$ in $m(t,r)$, which converts to the delta function.
Therefore it is computed as%
\footnote{ 
Here we present the computation of the delta function in the radial coordinate in a naive expression. This part can be computed in a more formally rigorous fashion by rewriting $\delta(r)$ in terms of the $d$-dimensional delta function $\delta^d(x)$ employing an equation obtained by evaluating the equation $\int d^d x\delta^d (x)=1$ with the polar coordinates. However such a formal description with extra complication might make physics obscure. Here we adopt a simple description to elucidate physics at the cost of the rigorous expression.
}
\beal{ 
M_2 
=& {d-2 \over 16\pi G_N} V_{d-2} m\int_0^\infty dr  (\delta(r) F(\frac{t + h(r)}\Delta)  )
= M F(\frac t\Delta ). 
}

The matter shell ends falling at $t=\Delta$. Afterwards the system becomes static, so is the energy flow. The energy configuration is obtained by taking the limit $t\to\Delta-0$. 
The total energy is $E=M$ as expected. 
Note that this result matches the ADM or quasi-local energy, which is computed in the asymptotic behavior of gravity in the system. 

\begin{figure}
 \begin{center}
  \includegraphics[scale=.5]{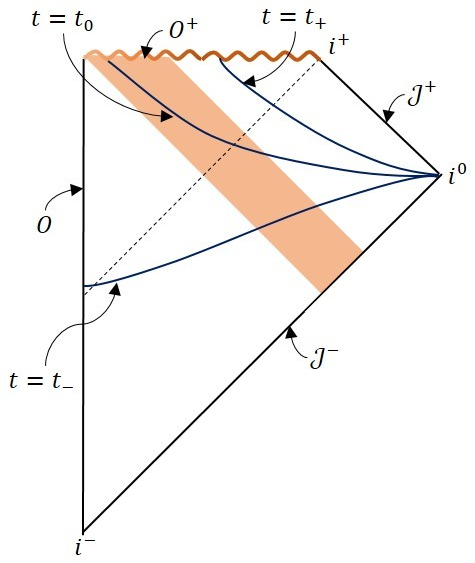}
 \end{center}\vspace{-0.5cm}
  \caption{The Carter-Penrose diagram of the spherical collapse of the thick light shell is depicted. We almost use conventional notation. $O$ is the origin, $O^+$ is the location of the curvature singularity, and the band in the middle describes the thick infalling light shell. The gradation of the color schematically describes the density of the energy. In addition three Eddington-Finkelstein time slices are schematically drawn, where $t_-<0<t_0 < \Delta < t_+$. Any intersecting point of the thick light shell with the time slice and that of the curvature singularity contributes to the integral of the total energy at the time. In particular the contribution from the curvature singularity is the mass of the black hole. 
  }
  \label{Fig:Energy}
\end{figure}

We summarize the result of energy flow in Table~\ref{EnergyConservation}. For instruction, we draw a picture of the Carter-Penrose diagram for the spherical collapse of thick light shell in Figure~\ref{Fig:Energy}.
\begin{table}[t]
 \label{}
 \begin{center}
  \begin{tabular}{|c|c|c|c|c|c|}
  \hline
 Time & $t$ & $t<0$ & $0\leq t<\Delta$ & $\Delta \leq t$  \\ 
  \hline
Energy of collapsing matter &  $M_1$  & $M$  & $M(1 - F(\frac{t}\Delta))$ & 0 \\
  \hline
Mass of black hole &  $M_2$  & 0   & $M F(\frac{t}\Delta) $ & $M$ \\
  \hline
Total energy &  $E$  & $M$ & $M$ &  $M$\\
  \hline
  \end{tabular}
 \caption{ The energy flow is shown. The total energy is conserved. }
 \label{EnergyConservation}
 \end{center}
\end{table}

We can also compute the energy inside the radius $R$ in the system denoted by $E(t,R)$:
\beal{
E(t,R)
=& -\int_{\sqrt{x^2}\leq R }\! d^{d-1} x\sqrt {|g|} T^t\!_t 
={d-2 \over 16\pi G_N} V_{d-2} \int_0^R dr  {\partial_r  (m(t,r) )} \nn
=& M F(\frac{t + h(R)}\Delta) .
}
This is a monotonically increasing function inside the fluid shells as long as $h(r)$ is. This can be used as an observable shell attribute. 

For concreteness we draw a picture of gravitational collapse of thick matter shell under pressure from the transverse direction without any shell crossing singularity in four dimensions. 
To this end we first determine the location of the final event horizons after the thick matter shell ends infalling. 
The determining equation of the event horizons is 
\be  
r^3 + \frac3\Lambda(-r+m) = 0. 
\label{eq:H}
\ee
This is solved by using the Cardano's formula as $r=r_k$, where 
\be 
r_k =\omega^k \sqrt[3]{-\frac{3m}{2\Lambda}+\sqrt{(\frac{3m}{2\Lambda})^2-(\frac 1\Lambda)^3}}+\omega^{3-k}\sqrt[3]{-\frac{3m}{2\Lambda}-\sqrt{(\frac{3m}{2\Lambda})^2-(\frac 1\Lambda)^3}}, ~~~~~
\ee
with $k=0,1,2$, $\omega={-1+\sqrt3\over2}$ the cube root of unity.
The discriminant is computed as $3(\frac3\Lambda )^2(-9m^2 + \frac4\Lambda )$.
Therefore for negative cosmological constant the discriminant is negative and only one positive solution exists as $r=r_0$. On the other hand, for positive cosmological constant smaller than $\Lambda_0=\frac{4}{9m^2}$ all three solutions for \eqref{eq:H} are real, two of which are positive, $r_1, r_2>0$, while for $\Lambda > \Lambda_0$ there is no positive real solution, so that the horizons do not exist \cite{Hayward:1993tt}. 
\begin{table}[thb]
 \label{}
 \begin{center}
  \begin{tabular}{|c|c|c|c|c|c|c|}
  \hline
Cosmological constant & $\Lambda$ & $\Lambda<0$ & $0$ & $0<\Lambda\leq\Lambda_0$ & $\Lambda_0<\Lambda$ \\ 
  \hline
Black hole horizon &  $r_{\rm BH}$  & $r_0$  & $m$ & $r_1$ & - \\
  \hline
Cosmological horizon &  $r_{\rm dS}$  & - & - & $r_2$ & - \\
  \hline
  \end{tabular}
 \caption{ The locations of the final black hole event horizon and the cosmological one after the end of the matter infalling are shown with the final mass of the black hole fixed. ``-'' means the absence of the corresponding horizon.  }
 \label{EnergyConservation}
 \end{center}
\end{table}
For positive cosmological constant, as the final mass of the black hole increases, the black hole event horizon radius $r_1$ increases while the cosmological one $r_2$ decreases \cite{Gibbons:1977mu}. 

\begin{figure}
 \begin{center}
  \includegraphics[scale=.5]{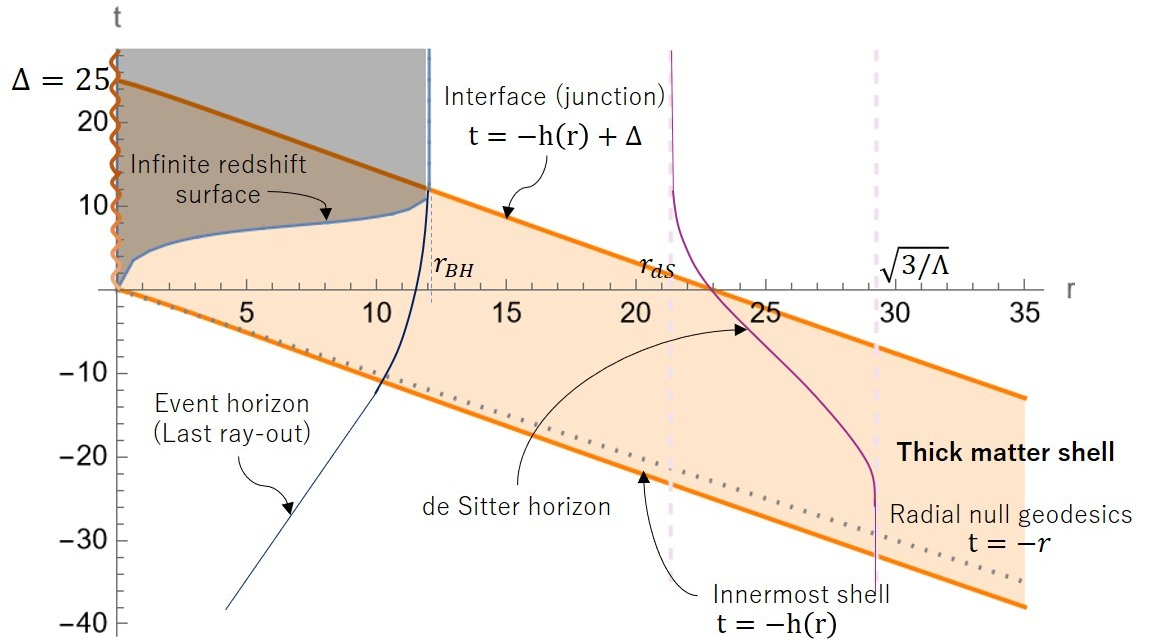}
 \end{center}\vspace{-0.5cm}
  \caption{A solution of the gravitational collapse of the thick matter shell for positive cosmological constant without any shell crossing singularity is depicted. The parameters are chosen such that $\bar v=0,\bar w=0.1, \Delta=25, m=10, \Lambda=0.035, \epsilon=0.05$, and an upslope function $F(x)=\sin^2{\pi x\over 2}$. For negative cosmological constant the corresponding picture is qualitatively the same except the disappearance of the de Sitter horizon. It can be seen that the infinite redshift surface monotonically increases and the flux energy condition \eqref{FEC0} is satisfied outside the emergent event horizon.} 
  \label{Fig:D25m10L0035}
\end{figure}

For concreteness we choose an upslope function as $F(x)=\sin^2{\pi x\over 2}$, whose inverse function is $F^{-1}(y)=\frac2\pi\arcsin\sqrt{y}$. 
Then the growing infinite redshift surface is 
\be 
t = -h(r) +\Delta F^{-1}(\frac{r}m(1-\frac\Lambda3r^2)), 
\ee
up to when the surface intersects the interface or the outermost thick matter shell. Afterwards it matches the final black hole horizon $r=r_{\rm BH}$.
The time for the surface to reach $r=r_{BH}$ is $t_{BH}=\Delta-h(r_{BH})$. 
The light-ray which passes through the intersecting point of the infinite redshift surface and the interface, $(r_{BH},t_{BH})$, does not go out, and any light-ray emitted after the last light-ray is trapped inside the horizon.  
Thus the orbit of the last light-ray describes the radius of the growing black hole event horizon, which is larger than the infinite redshift radius at an equal time until it reaches the final black hole horizon radius. The orbit of the last light-ray satisfies an integral equation $t =\int_{r_{BH}}^r {1-u \over 1+u} dr +t_{BH}$. In particular, before an outgoing light-ray reaches the thick matter shell, the orbit can be computed as $t=\frac{1+\frac{\Lambda r^2}{3} }{1-\frac{\Lambda r^2}{3}}(r-r_{BH})+t_{BH}$.
We draw a figure for a physically preferable solution in this setup in Fig.~\ref{Fig:D25m10L0035}.
 
\section{Discussion}
\label{Discussion}

We have investigated an analytic model of spherical gravitational collapse of thick matter shell without any radiation, where the metric is described by the Eddington-Finkelstein coordinate system with the mass parameter replaced with a function of the radial coordinate and the time.
We have argued a condition which makes a solution free from a shell crossing singularity, which was determined by requesting the monotonic infalling matter and the monotonic grow of infinite redshift surface in general dimensions.

For general analysis we have imposed an equation of state for the outermost matter consisting of the shell where pressure as a mixture of the longitudinal component and the transverse one is proportional to the energy density. 
Solving this equation of state at interface we have determined the form of the function describing the orbit of the matter shell or its derivative analytically, and discussed a parameter region where the solution is free from a shell crossing singularity. 
We have finally confirmed that employing the definition of the energy presented in \cite{Aoki:2020prb} the total energy of the system is invariant under the ordinary time translation.  

In order to avoid a shell crossing singularity for the infalling matter, we have imposed the function to describe the infalling matter as well as the emergent infinite redshift surface to monotonically increase. This condition is summarized as \eqref{NoCrush}. 
At three dimensions this simple inequality leads to a well-known result that negative cosmological constant is required for a black hole to exist. 
It would be interesting to investigate this inequality at four dimensions and extract any non-trivial condition for a black hole to be present. 

This paper has clarified that a gravitational collapse of matter shell generally suffers from shell crossing singularity and that even so it is possible to construct analytic solutions by choosing parameters suitably. 
This is an encouraging result to find such a solution describing gravitational collapse by specifying matter. 
The construction of such a solution with matter specified may give a hint to clarify the meaning of the off-diagonal component of the energy momentum tensor. As seen from the construction, the anisotropic form of fluid is important to construct the analytic solutions collapsing into a black hole. 
It would be interesting to understand the physical importance of the heat flux in this process. 

The system studied in this paper describes a rather simple gravitational collapse, but will be useful as a toy model.  
It is also possible to investigate this system by adding other structures keeping the spherical symmetry. One is to make the radius part of the metric dependent on time. In such a situation, the relative velocity of adjacent fluid element inside the same shell becomes non-trivial and will play an important role to construct a solution \cite{Misner:1964je}. Another interesting extension is to take into account the radiation and compute the energy of emitted gravitational wave. (See also  \cite{Alberghi:1998xe,Alberghi:2001cm,Baccetti:2016lsb}.) 
Although the definition of the energy used in this paper is based on the matter energy-momentum tensor, it could be possible to also compute the energy of the gravitational wave by computing the difference between the energy of the matter at the initial state and that of the final one. 
It would be of interest to compare the result to the one obtained by any other known approach. 

We hope to come back to these issues in near future. 

\section*{Acknowledgement}
The author would like to thank Sinya Aoki, Tetsuya Onogi for participation at an early stage of the project. 
He is also grateful to Tetsuya Shiromizu, Shigeki Sugimoto and especially Sinya Aoki for useful discussion and valuable comments on the draft.   
This work is supported in part by the Grant-in-Aid of the Japanese Ministry of Education, Sciences and Technology, Sports and Culture (MEXT) for Scientific Research (No.~JP19K03847). 

\paragraph{Data availability}
The data that support the findings of this study are available from the corresponding author upon reasonable request. 

\appendix
\section{Comment on standard energy conditions}
\label{EnergyCondition}

In the main text, we constrained a parametric function $m(t,r)$ or $h(r)$ to satisfy some physical conditions such as the absence of the shell crossing singularity, the monotonic increase of the emergent infinite redshift surface and some energy conditions. 
Here, for convenience to readers, we give a comment on standard energy conditions in the current context of gravitational collapse following \cite{Maeda:2018hqu}. (See also \cite{Curiel_2017,Martin-Moruno:2017exc} for reviews and references.) To be concrete and explicit, we first diagonalize the original Eddington-Finkelstein metric \eqref{EF} using its vielbein given by 
\be 
(E^{\hat\mu}\!_\nu) = 
\bordermatrix{
& t & r & j \cr
\hat t&  \sqrt{1+u} & \frac u{\sqrt{1+u}} & 0 \cr
\hat r& 0 &  \frac1{\sqrt{1+u}} & 0 \cr 
\hat i& 0 & 0 & r e^{\hat i}\!_j
},
\ee
where the Greek indices with the hat symbol represent the local Lorentz ones and $e^{\hat i}\!_j$ is the vielbein for the internal Einstein manifold. 
In this local Lorentz frame, the energy momentum tensor for infalling fluid takes the form of the type II classified in \cite{Hawking:1973uf} 
such that 
\beal{
(T_{\hat\mu\hat\nu}) = 
\pmat{
\varrho + \chi & \chi &0 \cr
\chi &  -\varrho+\chi & 0 \cr
0 &  0 &  \delta_{\hat i\hat j}p_j
},
\label{LLF1}
}
where 
\beal{
\varrho  
=&{d-2 \over 16\pi G_N } {(\partial_r- \partial_{t})m(t,r)\over r^{d-2}}  , ~~  
\chi
= {d-2 \over 16\pi G_N } \frac{ \partial_t m(t,r) }{r^{d-2} (1-\frac{2\Lambda r^2}{(d-1)(d-2)}  - {m (t,r)\over r^{d-3}})} , ~~ 
p_j
= p_\parallel
.
\label{LLF2}
}
Compared to the expression \eqref{EMT} in terms of the original Eddington-Finkelstein coordinate system, some components of energy momentum tensor containing $\chi$ suffer from the coordinate singularity located at infinite redshift surface, and macroscopic quantities read off from this frame as well.

Assuming the validity to discuss causality even in the presence of coordinate singularity in macroscopic quantities, we apply the results of some standard energy conditions to the current system. It is instructive here to investigate a so-called weak energy condition, which is described as $T_{\mu\nu}t^\mu t^\nu \geq 0$ for any timelike vector $t^\mu$. In the region outside the outer event horizon, which means the de-Sitter one if there is a positive cosmological constant, this condition reduces for the type II energy momentum tensor to $\varrho\geq0, \chi\geq0, \varrho+p_\parallel\geq0$ as computed in references. 
Using \eqref{LLF2}, we rewrite the first inequality $\varrho\geq0$ as $\partial_rm(t,r)\geq  \partial_{t}m(t,r)$, which is comparable to \eqref{rhoPositive} in the main text.  
In the shell ansatz \eqref{matterProfile}, these inequalities respectively become 
\bes{ 
& F'(x) \frac{h'(r)-1}\Delta \geq 0, ~
F'(x) \geq0, ~\\
& F'(x) \frac{h'(r)-1}\Delta  \geq \frac{r}{d-2} \( {F''(x) \over \Delta^2}  ( h'(r)^2 -2h'(r) + 1 ) + \frac{ F'(x)}\Delta h''(r) \) , 
\label{WEC}
}
where $x=\dfrac{t+h(r)}{\Delta}$. The first two inequalities imply that $h'(r) \geq 1$, which is formally comparable to the flux energy condition \eqref{FEC0} in the main text. Note that a causal vector considered in the main text is a radial one, so there is no analog of the 3rd inequality in \eqref{WEC} in the main text.

\bibliographystyle{utphys}
\bibliography{collapse}

\end{document}